# Heterogeneous Singlet Fission in a Covalently Linked Pentacene Dimer

Woojae Kim[1,5,†], Naitik A. Panjwani[3,†,*], K. C. Krishnapriya[2], Kanad Majumder[2], Jyotishman Dasgupta[4], Robert Bittl[3], Satish Patil[2,*], Andrew J. Musser[1,*]

[1]Department of Chemistry and Chemical Biology, Cornell University, Ithaca, New York 14853, USA

[2]Solid State and Structural Chemistry Unit, Indian Institute of Science, Bangalore 560012, India

[3]Berlin Joint EPR Lab, Fachbereich Physik, Freie Universität Berlin, 14195 Berlin, Germany

[4]Department of Chemical Sciences, Tata Institute of Fundamental Research, Mumbai 400005, India

[5]Department of Chemistry, Yonsei University, Seoul 03722, Republic of Korea

[†]These authors contributed equally: Woojae Kim, Naitik A. Panjwani

Correspondence to: npanj@zedat.fu-berlin.de, spatil@iisc.ac.in, ajm557@cornell.edu

# Abstract

Molecular dimers are widely utilized as a tool to investigate the structure-property relationships behind the complex photophysical processes of condensed-phase systems, where structural tuning remains a challenge. This approach often implicitly treats the dimers as 'static', with their relevant state energies and couplings determined by their optimized geometry. Here, we consider the shortcomings of this approach–dimers are more accurately treated as 'dynamic' model systems, with the potential for significant conformational heterogeneity that evolves in time and is intimately connected with interchromophore coupling strengths. We highlight this concept in the singlet fission dynamics of a pentacene dimer that is covalently linked through phenyl-diketopyrrolopyrrole and acetylene bridges. Unrestricted rotations around the acetylene bridges lead to a vast array of rotational conformers in the ground state. Consequently, we find that every step in the cascade of singlet fission processes– triplet-pair formation from the initial $S_1$ state, triplet-pair recombination, spin evolution within the triplet pair, and free triplet formation–is qualitatively and quantitatively altered by the conformer geometry. Through systematic excitation-energy-dependent transient absorption experiments at room temperature, we find evidence of dynamic interconversion between conformers on the multiple TT surfaces. Further experiments conducted using transient electron paramagnetic resonance spectroscopy on a frozen solution at 150 K emphasizes the significance of static disorder in a dimer. The results reveal the presence of conformational sub-populations which leads to excitation dependent quintet spectra, with increasing quintet and triplet intensities with higher energy excitation. These phenomena demand consideration of multidimensional potential energy surfaces that define multiple sub-ensembles in the excited state, a picture we refer to as 'heterogeneous singlet fission'. More broadly, these results call into question the general 'static' approach with a single sub-ensemble to molecular dimer photophysics, that each step in consecutive excited-state relaxation pathways can be delineated with a single, unique rate constant and yield.

# Introduction

The bulk of optoelectronic processes in molecular systems are intermolecular in nature.[1–3] Phenomena such as energy transfer, charge transport, electron-hole separation, excimer formation, and exciton multiplication through singlet fission (SF) are chiefly studied in solid films,[4–9] which are most relevant for eventual applications. However, it remains an outstanding challenge to rationally control the intermolecular interactions that drive these processes in the solid state. Moreover, the substantial disorder present in typical thin films can result in a wide distribution of underlying photophysical rates, which can exhibit profound differences from the corresponding single crystals.[10–12] One of the most effective solutions to this challenge is the use of tailored molecular dimers as model systems for complex solid-phase photophysics.[13–15] In dilute solution, the photophysical interactions are attributed purely to well-controlled through-bond and through-space couplings.[16–18] Due to the structural simplicity of these models, this strategy has yielded powerful mechanistic insight into the structure-property relationships behind excited-state dynamics from ultrafast charge separation and SF to the slow evolution of entangled spin states.[19–24] And yet, some caution is warranted with this approach. Though they often exhibit narrower linewidths and thus less disorder than thin films, such dimers can exhibit a broad array of interchromophore orientations, depending on the chemical nature and position of the linkers.[25–27] This intramolecular conformational heterogeneity is not often treated as an important contributor to either the dynamics or the balance of product states. Instead, common practice is to consider only a single ensemble, assuming quasi-static structures based on the equilibrium geometries of the relevant electronic states, for the interpretation of the photophysics in dimers.

This kind of interpretation is widespread in the study of SF, in which a photoexcited singlet exciton ($S_1$) generates two triplet excitons ($T_1+T_1$) through a triplet-pair intermediate (TT).[28–32] Both in thin-film studies and dimer-based intramolecular work, the initial formation of TT and its subsequent dissociation into $T_1+T_1$ or recombination to the ground state are chiefly modeled with single-exponential kinetics.[8,17] This approach signifies that a system's rates of TT formation/recombination/dissociation can be pinpointed to single, unique values. Accordingly, reports of new SF materials describe 'the' rate constant of TT formation or separation that it exhibits. In dimer studies, this framework has highlighted important mechanistic trends: with noteworthy exceptions,[33] a molecule showing fast TT formation also exhibits fast TT recombination.[17] But moving beyond general dynamical trends to map the rate constants onto specific couplings—an important step for detailed structural design rules—has proved challenging, as in highly rigid norbornyl-linked dimers.[34] There is growing awareness of the limitations of the standard 'static' framework in non-rigid systems, which can exhibit multiple important conformations. For instance, a pentacene dimer connected through a flexible crown ether linker revealed solvent-controlled formation of two distinct sub-ensembles in the ground state (weakly coupled monomer-like, vs. strongly coupled aggregate), with corresponding TT formation rates that differed by a factor of 250.[35] Similar biexponential TT formation dynamics were reported in π-stacked pentacene dimers encapsulated by polyaromatic capsules, interpreted as an effect of different stacking conformations.[36] Furthermore, a few reports have hinted at the presence of multiple sub-ensembles in conformationally flexible molecular dimers by controlling the viscosity of the environment through polymer matrices or temperature, revealing significant changes in TT formation and decay dynamics as different sub-ensembles are excited.[37,38] Given this increasing evidence of heterogeneity in the ground/excited states and structural dynamics between different conformers, an important question moving forward is how to properly capture its impact on the dynamics and products of the SF pathway.

In this study, we revisit the SF dynamics of a TIPS-pentacene (P-TIPS) dimer bridged via acetylene linkers at the 6,6'-position with a phenyl-diketopyrrolopyrrole (PDPP) (Figure 1A). We previously reported efficient and fast intramolecular SF in this dimer enabled by synthetic control of the spin density localization and by tuning the electronic nature of the bridge.[39] Here, we focus instead on the role of the acetylene linkers in enabling nearly unimpeded rotation about its axis, which induces conformational heterogeneity with various dihedral angles among the two P-TIPS units and the PDPP bridge.[37,38] This gives rise to different strengths of inter-pentacene electronic coupling and thereby affects the rate constants for the SF dynamics. Through systematic excitation-energy-dependent transient absorption (TA) experiments at room temperature, we demonstrate that intramolecular $^1$(TT) formation and recombination cannot simply be defined with monoexponential kinetics and can be markedly more complex than what is reported in many systems to date. We suggest a dynamic picture of intramolecular SF, based on multiple sub-ensembles in the ground state and conversion between these in the excited state. We find that the full range of SF dynamics, spanning from 'ultrafast' (<200 femtosecond) 'to 'slow' (<10 ps) TT formation and including multiple TT recombination channels, are controlled by photoexcitation of specific sub-ensembles with different excitation energy. Strikingly, we find that these excitation-energy effects persist out to microsecond spin evolution processes, resulting in unprecedented selectivity in the spin polarization patterns of $^5$(TT) states observed in transient electron paramagnetic resonance spectroscopy. Our findings present a cautionary note that the perspective of both static and dynamic conformational heterogeneity is necessary to describe the photophysics of even simple molecular dimers. Though it poses challenges, embracing this approach offers the scope for ever deeper insight into the structure-property relationships behind SF and provides a unique avenue to steer the outcome of photophysical processes.

# Results

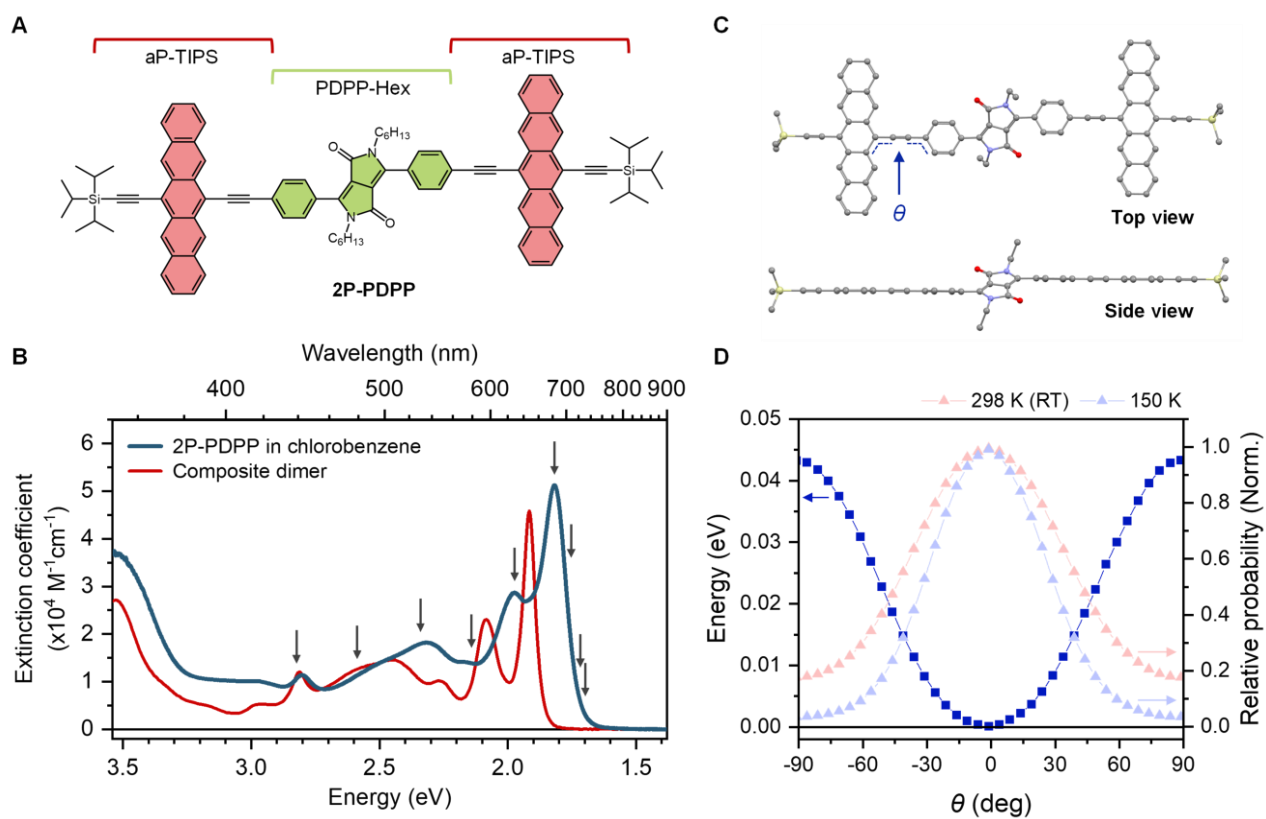

**Figure 1.** (A) Chemical structure of 2P-PDPP composed of two aP-TIPS units linked by a PDPP-Hex chromophore. (B) Molar extinction coefficient spectrum of 2P-PDPP in chlorobenzene (Solid blue line). The solid red line is a composite spectrum assuming the absence of electronic interactions between aP-TIPS and PDPP-Hex units. The arrows denote excitation energies used in the TA experiments. (C) The energy-minimized structure of 2P-PDPP (CAM-B3LYP/6-31g(d) level) in the ground state. Hydrogens are omitted for clarity. (D) Calculated energy curve (rectangle) in the ground state of 2P-PDPP along the torsional coordinate $\theta$, i.e., the dihedral angle between phenyl and pentacene planes. Normalized Boltzmann distributions (triangle) in two different temperatures (light red – 298 K, light blue – 150 K) are also shown.

**Basic Photophysics and Molecular Structure.** As we previously reported, 2P-PDPP is composed of two aP-TIPS chromophores (here, 'a' denotes asymmetric because it has only one TIPS substituent unlike P-TIPS, Figure S1), which are linked by a PDPP-Hex bridge through ethynylene units attached at the 6 or 6' position of each pentacene.[39] We observe a clear mismatch between the steady-state absorption spectrum of 2P-PDPP in chlorobenzene and a composite spectrum assuming non-interacting constituents (Figure 1B and Figure S1). Most importantly, the signature vibronic bands of the parent chromophores are all considerably red-shifted and broadened. The former we attribute to strong electronic interactions between aP-TIPS units via the PDPP-Hex linker, despite their very long center-to-center distance of 23 Å. Indeed, this red-shift is markedly more prominent than in other pentacene dimers, for example, the 2,2'- or 6,6'-directly-linked dimers,[13,15] despite their much shorter center-to-center distances. This result highlights the impact of fully conjugated linkers. As for the broadening effect, contrary to the Lorentzian-like line shape of bare P-TIPS, 2P-PDPP instead follows a Gaussian-like line shape with significantly increased bandwidth, suggesting inhomogeneous broadening. We attribute this effect to conformational heterogeneity due to the shallow torsional potential energy curve along the rotation axis of the ethynylene units, as confirmed by DFT calculations (Figure 1D). The relatively unimpeded rotation of this bond results in a broad distribution of conformers with different dihedral angles between the two aP-TIPS units and the PDPP-Hex bridge even at 150 K and, consequently, varying degrees of conjugation and electronic coupling strength.[38]

**Heterogeneous Singlet Fission.** Figure 2 shows transient absorption spectra of 2P-PDPP in chlorobenzene after photoexcitation at 1.97 eV corresponding to the 0-1 vibronic absorption. We probed the transient response from 3.44 to 0.75 eV, covering a wide range from ultraviolet and visible to near-infrared regions (See the Methods section for experimental details). Right after photoexcitation (0.3 ps), positive bands around 1.9 eV closely match the ground-state absorption to the $S_1$ state and can be assigned to ground-state bleaching (GSB). A distinct negative photoinduced absorption (PIA) band at 0.93 eV is captured in the near-infrared, and several further PIA bands are revealed in the ultraviolet and visible region (3.44 to 2.20 eV). Interestingly, through the course of the spectral evolution up to 5 ps, we observe five isosbestic points offset from the baseline (black arrows in Figure 2), a clear marker of population transfer dynamics between two different excited states. Based on our previous work and other reports dealing with the excited-state dynamics of pentacene dimers,[13–15,33–35,39–42] the observed spectral evolution can be ascribed to $^1$(TT) formation from the initially populated $S_1$ state. Furthermore, we note that even at the earliest time delay beyond our instrument response (300 fs), we can already distinguish the characteristic spectral signatures of $^1$(TT), namely vibronic PIA bands near 2.48 eV and an additional band around 3.2 eV.[13–15,35] The rapid formation of so much $^1$(TT) population suggests the possibility of an ultrafast, coherent triplet pair formation channel in 2P-PDPP despite the very long pentacene center-to-center distance (23 Å, Figure 1C). We return to this topic below. Beyond 5 ps, the principal transient signals from $^1$(TT) decay uniformly up to 2 ns, leaving a small residual population that does not decay within the detection limit of our delay stage (7.7 ns).

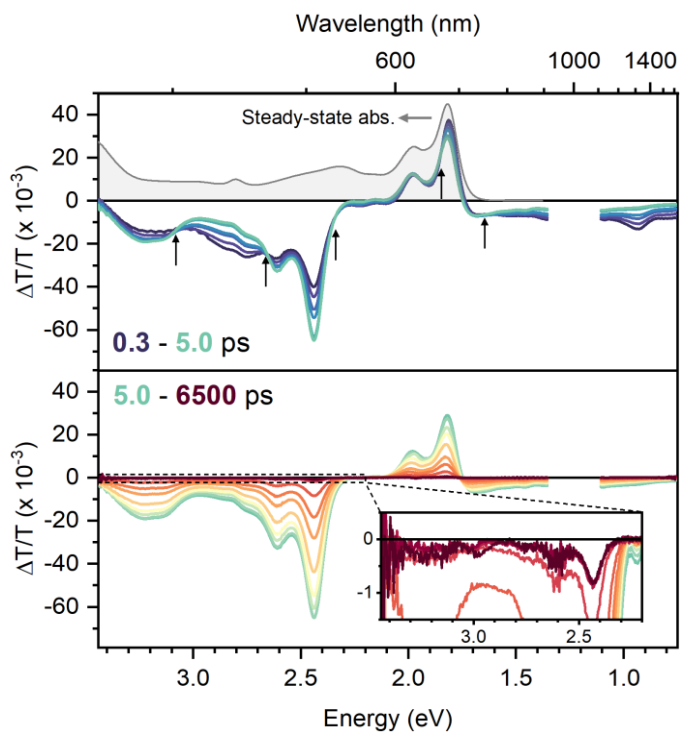

**Figure 2.** TA spectra of 2P-PDPP in chlorobenzene at short (0.3 - 5.0 ps, top) and long (5.0 - 6500 ps, bottom) delay times after photoexcitation at 1.97 eV (53.1 µJ/cm$^2$). The steady-state absorption spectrum (grey shaded) is shown in the upper panel to show the contribution of GSB. The black arrows in the upper panel indicate isosbestic points offset from the baseline, implying direct population transfer from the S$_1$ to the $^1$(TT) states. The inset in the lower panel highlights the existence of the long-lived PIA signals after 2 ns.

To investigate the possible contribution of conformational heterogeneity, we performed the same measurements for the full range of excitation energies indicated in Figure 1B. In all cases, we observe qualitatively similar TA spectral evolution (Figures S2 and S3), but the corresponding kinetics are systematically modulated. Figure 2A shows normalized ΔT/T kinetics of the main $^1$(TT) absorption band (averaged over 2.53-2.38 eV). As excitation energy is reduced, both the formation and recombination of $^1$(TT) accelerate, and the intensity of the long-lived component decreases. We quantify these trends with a global exponential fitting procedure. We find that the kinetics of $^1$(TT) formation and $^1$(TT) recombination do not follow simple monoexponential functions as previously reported.[39] Instead, a minimum of five decay components are required to capture the full dynamics of the system and minimize residuals (see Figures S4-S7 for analysis of the appropriateness of this model). This model entails two rise components of 0.3 and 2.3 ps, two decay components of 72 and 234 ps, and an additional non-decaying component $\gg$7 ns (Figure 2B).

This result implies that the SF dynamics of 2P-PDPP are substantially more complicated than what is typically reported of other pentacene dimers. In an isolated measurement, it would be tempting to assign these dynamics to multi-step relaxation, $^1$(TT) formation, and subsequent recombination of a single ensemble. However, such a model would predict very specific forms for the excitation energy dependence. Vibrational relaxation, vibrational cooling, and/or solvation processes following the generation of $S_1$ with excess energy typically occur on timescales of a few hundreds of femtoseconds to a few tens of picoseconds, similar to common SF rates.[43,44] In TA measurements, these processes generally manifest as time-dependent blue shifts of the PIA bands.[45] This behavior is completely absent in data on 2P-PDPP regardless of the excitation energy, so we rule out significant contributions from them (Figure S8). Optically activated branching between different decay pathways has been reported in conjugated polymers and tetracene dimers.[22,46] However, we observe no major changes in the spectral shapes detected across our excitation series. That is, there are no distinct electronic states formed with excess energy, and the same qualitative photophysical pathway is accessed despite the >1 eV variation in excitation energy. Accordingly, we discard the possibility of optically activated branching and with it the single-ensemble picture.

Instead, we invoke the conformational disorder of 2P-PDPP along the triple bond inferred from the ground-state absorption and, with it, significant variation of the inter-pentacene electronic coupling strength. Our data strongly suggest there are multiple sub-populations in the excited state and, based on our observed biexponential rise and decay dynamics, we propose to map the population onto two representative ensembles. Specifically, we expect one ensemble to exhibit near-planar geometry with the strongest inter-pentacene coupling. Following the well-known positive correlation between $^1$(TT) formation and recombination rates in synthetically controlled dimers,[17] this population exhibits faster $^1$(TT) formation (0.3 ps) and recombination (72 ps). We denote this ensemble as population 'A', and its electronic states will be defined as $(S_1)_A$ and $^1(TT)_A$ below. On the other hand, the other population captures geometries with more torsional distortion between the aP-TIPS units and the PDPP-Hex bridge, with characteristically weaker electronic coupling. Accordingly, the $^1$(TT) formation (2.3 ps) and recombination (234 ps) dynamics are relatively slow. This we denote as population 'B', with states $(S_1)_B$ and $^1(TT)_B$.

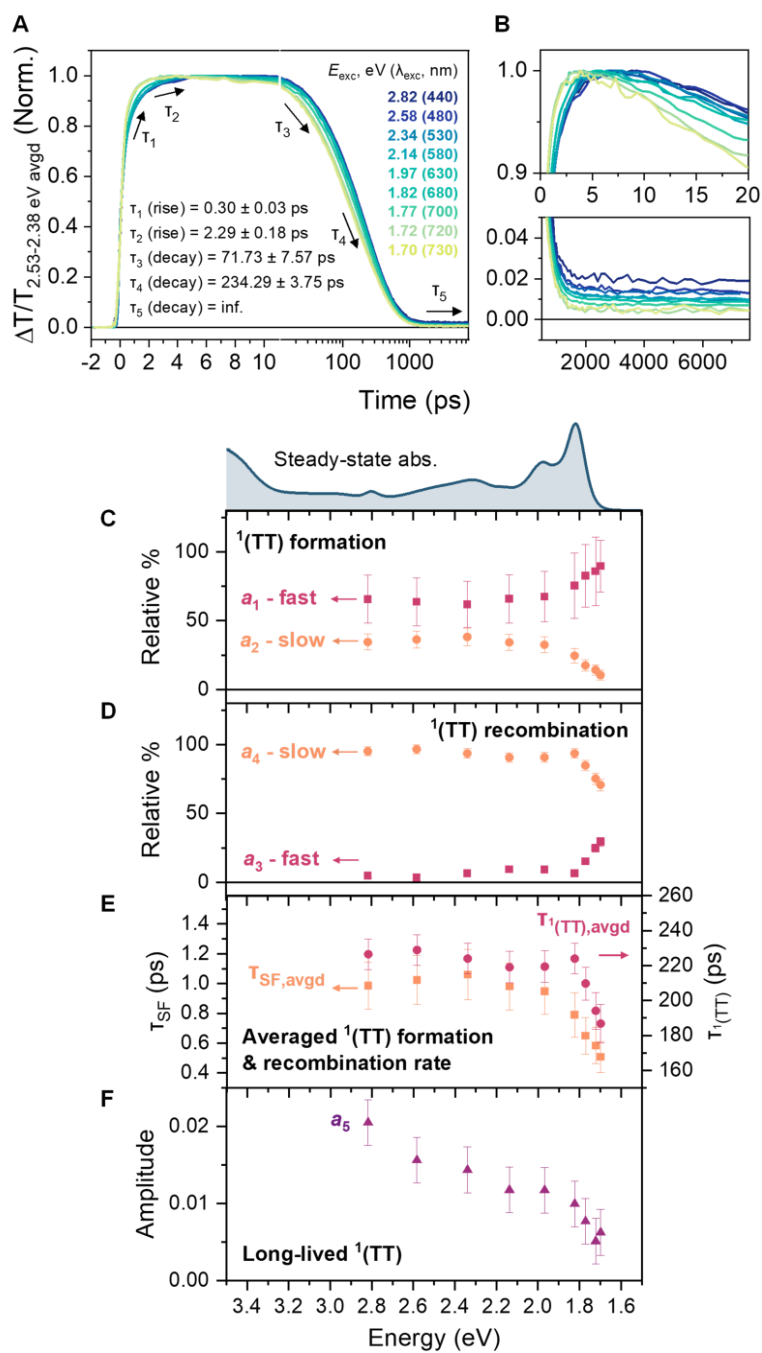

**Figure 3.** (A) Excitation energy-dependent normalized ΔT/T kinetics in $^1$(TT) absorption region (averaged from 2.53-2.38 eV, left). (B) The top and bottom panels show enlarged views of early-time (0-20 ps) and long-time (500-7600 ps) kinetics. All kinetics were globally fitted with a shared set of time constants; fit lines and residuals are shown in Supplementary Information. Global fit results: (C) Excitation energy dependence of the percentage contribution of fast ($\tau_1$ = 0.3 ps) versus slow ($\tau_2$ = 2.3 ps) components to the biexponential $^1$(TT) formation. (D) Excitation energy dependence of the percentage contribution of fast ($\tau_3$ = 72 ps) versus slow ($\tau_4$ = 234 ps) components to the biexponential $^1$(TT) recombination process. (E) Excitation energy dependence of the weighted-average time constants for $^1$TT formation and recombination. (F) Excitation energy dependence of the relative amplitude of the long-lived components ($\tau_5$ = inf.). The steady-state absorption spectrum is shown on top of panel C for reference.

By comparing the decay-associated amplitudes estimated from the global fit, we find that the relative contribution of the two sub-ensembles – and thus, presumably, their population – can be systematically controlled by the excitation energy (Figure 3C-D). In the range of excess-energy excitation (2.9-1.9 eV), we find that the biexponential $^1$(TT) formation kinetic is roughly constant, with weights of 65% for $\tau_{SF,Fast}$=0.3 ps ($(S_1)_A$ to $^1(TT)_A$) and 35% for $\tau_{SF,Slow}$=2.3 ps ($(S_1)_B$ to $^1(TT)_B$). Below 1.9 eV excitation, the contribution of $\tau_{SF,Fast}$ progressively becomes much larger than that of $\tau_{SF,Slow}$. We observe a qualitatively similar excitation energy dependence in the biexponential $^1$(TT) recombination kinetics, namely relatively constant behavior in the excess-energy range and a sharp uptick in the weight of $\tau_{TT,Fast}$ ($^1(TT)_A$ to the ground state) near band-edge excitation. From their strikingly similar excitation energy dependence, it is reasonable to assign the $\tau_{TT,Fast}$ components to the same excited population, and likewise for $\tau_{TT,Slow}$ ($^1(TT)_B$ to the ground state). We stress that, as in most SF dimers with well-spaced chromophores, there are no resolvable spectral signatures to distinguish the more-coupled from less-coupled triplet pairs. These populations are thus identified solely by their characteristic lifetimes.

These effects vary most strongly between band-edge excitation and 1.9 eV, which matches the expected absorption maximum of a non-interacting dimer (Figure 1B). This behavior is consistent with our sub-ensemble interpretation, where the excitation selects a variable balance between planar and distorted populations. However, we highlight that the balance between fast (A-type) and slow (B-type) components is reversed between the early-time (Figure 3C) and intermediate (Figure 3D) decay ranges. The fast-decaying form $^1(TT)_A$ converts nearly quantitatively to slow-decaying $^1(TT)_B$ under excess-energy conditions, but the conversion is partially suppressed with band-edge excitation. We return to these conformational dynamics in the Discussion below.

From the amplitude-weighted $\tau_{Fast}$ and $\tau_{Slow}$ components, we determine the excitation-energy-dependent averaged $^1(TT)$ formation and $^1(TT)$ recombination time constants, as shown in Figure 3E. These dynamics show qualitatively similar activation behavior, confirming the effects arise from the same states. A comparable effect is observed in the contribution of the long-lived state in the transient kinetics (Figure 3F), with the notable exception of a more pronounced rise in yield at high excitation energies. As discussed below, this long-lived state can be assigned to yet another triplet-pair sub-ensemble, $^1(TT)_C$, which we link to further spin evolution.

**Evidence of Conformational Disorder.** We further substantiate our model of multiple sub-ensembles by analysis of the excitation-energy-dependent TA spectra at early delay time (300 fs). Figures 4A and 4B show normalized TA spectra in the GSB and PIA regions. In general, GSB reflects the ground-state absorption spectrum of the photoexcited molecules and its shape and peak positions should be in line with the steady-state absorption, though often it can be distorted by overlapping PIA bands. For 2P-PDPP, we see such a match following excess energy excitation. On the other hand, as the excitation energy is reduced, the GSB 0-0 band peaks are systematically red-shifted. The red-shift is coincident with the onset of population 'A' dynamics, characteristic of planar conformers with enhanced inter-pentacene interactions. At the same time, we observe systematic shifts in the PIA bands in the NIR region (Figure 4B). The main $S_1$ PIA band is located at around 0.925 eV following high-energy excitation, whereas it reveals a blue shift to 1 eV and an increase in the intensity in the vicinity of the 0.8 eV region upon lowering the excitation energy. This trend is consistent with the gradual variation of the electronic structure along the torsional coordinate in 2P-PDPP, leading to stabilized $S_1$ states in more planar conformers. These effects demonstrate that the inhomogeneous broadening highlighted above reflects a range of separately addressable sub-ensembles, with different electronic structure defined by torsional disorder.

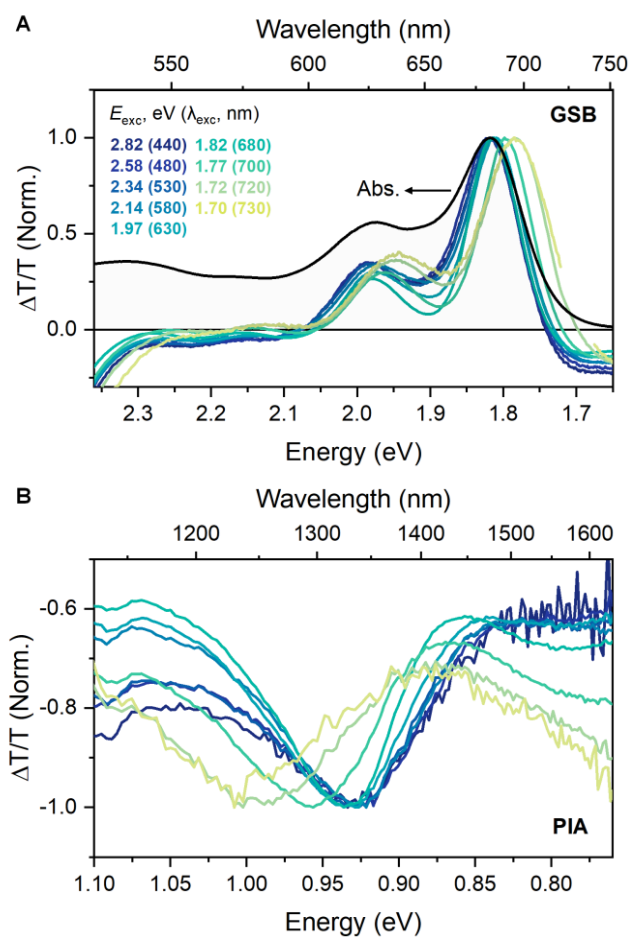

**Figure 4.** Excitation energy dependence of normalized ΔT/T spectra at around 300 fs in (A) GSB and (B) PIA regions. Steady-state absorption spectrum (black solid line) is also shown for comparison in panel A.

**Ultrafast Triplet Pair Formation.** The type of conformer excited has pronounced effects even on the earliest timescales detected. As noted above, 2P-PDPP reveals $^1$(TT) absorption bands at 300 fs delay time, implying an ultrafast SF channel to produce an appreciable $^1$(TT) population within our ~200 fs instrument response. Because of this initial mixed population, we cannot robustly deconvolve these initial dynamics using standard methods like global target analysis. Instead, we recall the multiple isosbestic points in the TA spectra detected on longer timescales (Figure 2). Their presence indicates that SF in 2P-PDPP can be regarded as a nonadiabatic population transfer from $S_1$ to $^1$(TT), rather than an adiabatic evolution along a single surface from $S_1$-like to $^1$(TT)-like character. We thus extract the buried TA signature of pure $S_1$ using decomposition of two TA spectra: the 0.5 ps spectrum, containing information on a mixture of $S_1$ and $^1$(TT), and the spectrum at 10 ps, where SF is complete and only $^1$(TT) is present. As shown in Figure 5A, the estimated $S_1$ spectrum shows the main PIA at about 2.8 eV, and the overall shape (peak positions and their ratios) closely matches the $S_1$ signatures in other pentacene dimers and P-TIPS (Figure S12).[34,38,40] This analysis permits the identification of another distinctive excitation-energy effect. Examining the 0.5 ps TA spectra at each excitation energy in Figure 5B, we observe a systematic reduction in the relative weight of the $S_1$ PIA in the region 2.6 eV to 3.3 eV as the excitation energy is reduced (The black arrow). We conclude that excitation of more planar conformers shifts the balance between $S_1$ and $^1$(TT) towards $^1$(TT) even on ultrafast timescales. We find that the ultrafast channel in 2P-PDPP can lead to approximately 30% $^1$(TT) yield within 300 fs (Figure S13). The detailed mechanism of this ultrafast process is beyond the scope of the current work and is under further investigation.

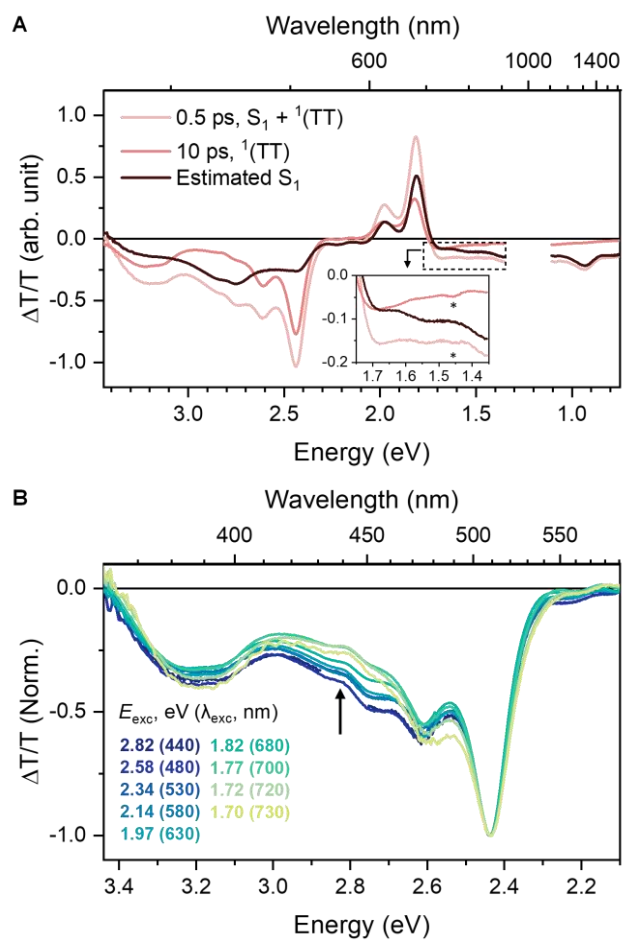

**Figure 5.** (A) Estimation of the pure $S_1$ TA spectrum by subtracting the 10 ps spectrum from the 0.5 ps spectrum. The 0.5 ps spectrum is first normalized to the PIA maximum. The 10 ps spectrum is scaled until the characteristic the $^1$(TT) absorption signatures in the estimated $S_1$ spectrum are minimized vanished. As an example, the inset highlights the $^1$(TT) PIA band at around 1.46 eV (asterisk), which is totally absent in the estimated pure $S_1$ spectrum. Spectral decomposition was performed using TA data following 1.97 eV excitation. (B) Excitation energy dependence of normalized PIA lineshape at 0.5 ps time delay, showing that the contribution of $S_1$ in the early time TA spectra becomes smaller as the excitation energy decreases (black arrow).

**The Nature of the Long-Lived Spin States.** In order to establish the fate of the long-lived $^1$(TT)$_C$ population, we performed transient electron paramagnetic resonance (trEPR) spectroscopy on a frozen 100 µM solution in a 1:1 toluene:*ortho*-dichlorobenzene solvent mixture, which has comparable dielectric properties to chlorobenzene used in the TA, at 150 K. We measure trEPR spectra at a range of excitation energies from 1.82 to 2.82 eV and find that at early delay after flash (DAF) time intervals (0.3-0.5 µs), the spectra are dominated with features corresponding to the quintet state $^5$(TT) (Figure 6A – transitions shown by labels A-E). The presence of $^5$(TT) categorically proves SF to be active at all excitation energies.

The quintet spectrum formed after excitation at all measured excitation energies shows the electron spin polarization (ESP) pattern of *aae*/*aea* (where *a* is enhanced absorption and *e* is emission) and can be simulated using the MATLAB toolbox EasySpin with zero-field splitting parameters (ZFS) $D_Q =$ 376.7 MHz, $E_Q = 12.7$ MHz and an exchange coupling (*J*) of 15 GHz.[47] Comparison of the spectral intensities at magnetic field positions 329.5 mT and 336.1 mT (Positions A and B in Figure 6A), associated with the Q$_{-2}$→Q$_{-1}$ and shoulder (canonical Z orientation) of Q$_0$↔Q$_{\pm1}$ transition, to the most absorptive quintet peak at 352.3 mT (Position D) show important differences. We observe no variation in the quintet dynamics (Figure S23), hence the spectral differences must reflect changes in the nature of the population (See SI trEPR section for further discussion). Simulations show that these differences in the spectral shape can be modelled with a decreasing Q$_{-2}$ and simultaneously increasing $Q_{-1}$ relative sublevel population as the excitation energy increases (See Figure 6A and 6C). At 0-0 band excitation (1.82 eV) the relative sublevel populations are [0.56, 0.07, 0.37, 0, 0] going from $Q_{-2}$ to $Q_{+2}$, while at high energy excitation (2.82 eV) the relative populations needed to simulate the experimental results are [0.47, 0.17, 0.36, 0, 0].

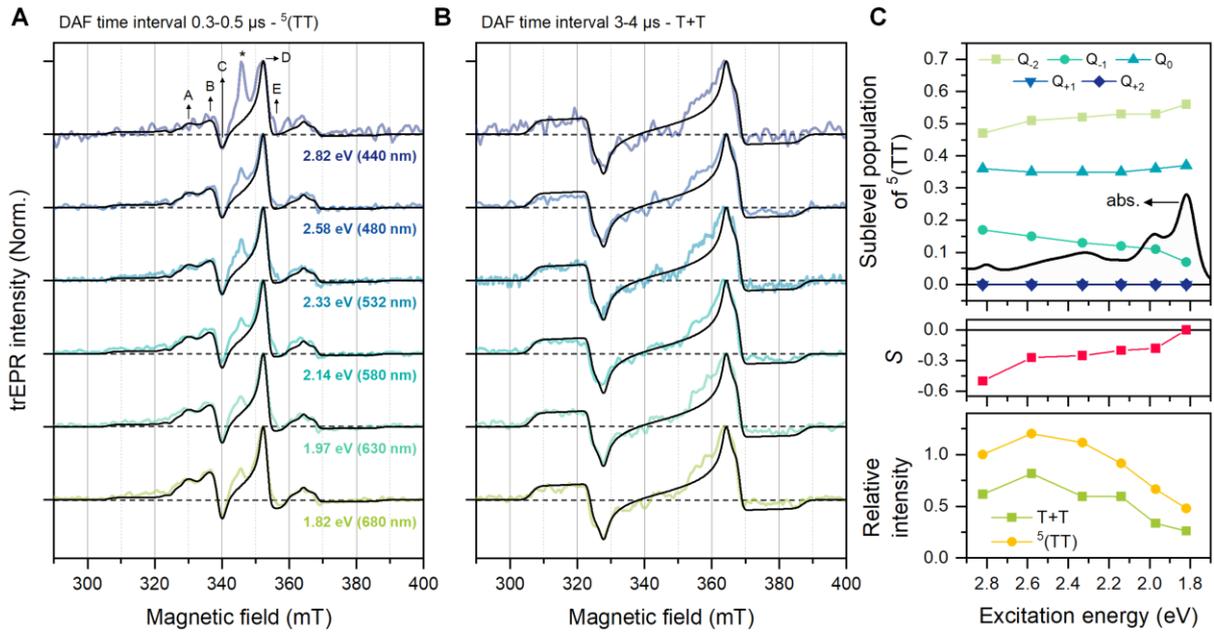

**Figure 6.** trEPR spectra at (A) early (0.3-0.5 µs) and (B) late (3-4 µs) delay after flash time intervals after excitation at different excitation energies. Panels A and B show the contribution of $^5$(TT), and free triplets (T+T) formed via TT dissociation, respectively. The asterisk in panel A indicates the $g$=2 radical signal. (C) Relative sublevel population of $^5$(TT) (top), orientational ordering parameter $S$ (middle), and relative intensities of $^5$(TT) and T+T, depending on excitation energies. The steady-state absorption spectrum in the top panel is also shown for reference.

At the same time, we find an increase in the orientational ordering parameter ($S$), which accounts for the degree of selectivity in the spin polarization with the orientation of the molecular axes relative to the external magnetic field. We see that an increasingly negative $S$ is needed to simulate the experimental data when using higher excitation energies ( $S_{1.82\,\text{eV}} = 0$, $S_{1.97\,\text{eV}} = -0.18$, $S_{2.14\,\text{eV}} = -0.2$, $S_{2.33\,\text{eV}} = -0.25$, $S_{2.58\,\text{eV}} = -0.27$, $S_{2.82\,\text{eV}} = -0.5$). The increasingly negative $S$ results in decreased intensity at positions B and E compared to D and C. Further discussion related to the orientational ordering parameter are presented in SI trEPR Section. We expect such orientation selectivity to arise due to dipolar contributions to spin mixing between the $^1$(TT) and $^5$(TT) manifolds.

It is important to note that while there are significant changes to the quintet spectrum as a function of excitation energy, the resonance position of the different transitions within the quintet manifold do not change. The separation between the most intense absorptive (Position D) and emissive (Position C) peaks of the $Q_0 \leftrightarrow Q_{\pm 1}$ transition is given by $|D_Q| - 3|E_Q|$ and when the two chromophores involved in SF are co-planar $|D_Q| - 3|E_Q| = \frac{1}{3}(|D_T| - 3|E_T|)$, where $D$ and $E$ correspond to the quintet and triplet ZFS parameters. We find that the separation between the most absorptive and emissive peaks of the $Q_0 \leftrightarrow Q_{\pm 1}$ transitions in our spectra matches the 1/3 relationship with the free triplet ZFS parameters. Hence, in the dimer molecules forming long-lived states, the two aP-TIPS moieties must be in an approximately co-planar configuration. Any deviation from this co-planarity would lead to a decreased quintet spectral width.

Along with the changing quintet spectrum, we also observe an increase in the *g*=2 signal centered at 346.5 mT (Figure 6A, position marked with an asterisk) which appears as a purely absorptive feature and is largest at the highest excitation energy of 2.82 eV. This absorptive feature increases slightly going from excitation at 1.82 to 2.58 eV and considerably more when excitation is at 2.82 eV. The ground state absorption band near 2.82 eV is known to have significant intramolecular charge transfer (CT) character.[23,48] While we do not resolve enough structure to unequivocally attribute the *g*=2 signal to a CT state, its narrow lineshape and the nature of the 2.82 eV excitation band together suggest this is the most likely origin. Further investigation is warranted to establish the origin and its role in the SF process.

The quintet and *g*=2 signal both show decay on a similar time scale whereby at a DAF time interval of 3-4 μs both features disappear (Figure 6B). At these later times, we observe a broader trEPR signal with an ESP of *aee*/*aae* pattern corresponding to the SF-born free triplet (T$_1$ + T$_1$).[24,49,50] Unlike the quintet, the ESP of the free triplet signal shows no significant differences at the various excitation energies, suggesting the mechanism behind the spectral differences only significantly affects the coupled TT pair state and is lost when coherence between the triplets in the pair state dephases. The free triplet spectrum can be simulated using $D_T = 1130$ MHz, $E_T = 38$ MHz and a relative sublevel population of [0.25, 0.75, 0.00] from $T_{-1}$ to $T_{+1}$. The asymmetry in the free triplet signal arises from the greater population of the $T_{-1}$ relative to $T_{+1}$ sublevels, which in part is due to the significant $Q_{-2}$ population, as population in $Q_{-2}$ results in 2×$T_{-1}$.[24] Similarly, to the quintet dynamics, the triplet dynamics do not have any excitation energy dependency (Figure S23). We note that there is additional absorptive signal around 355 mT, which is not accounted for by the triplet simulation in Figure 6B, for all excitation energies. This likely corresponds to remaining quintet signatures with an inverted spin polarization compared to the early time spectra in Figure 6A.

Finally, we consider the relative intensities of the most absorptive peak for the quintet and free triplet signals both in the magnetic field and in time, as a function of excitation energy. Taking into account both the excitation fluence and extinction coefficients, we see that the trEPR intensity for quintets and free triplets rises when going from 1.82 eV to 2.58 eV before slightly decreasing again by 2.82 eV (Figure 6C). While trEPR is not quantitative, a relative comparison is reasonable and consistent with the TA results showing larger yields of long-lived states with higher state excitation (Figure 3F). Interestingly, in our solution-phase TA experiments the excitation energy dependence is linked to dynamic conformational interconversion, but this is not possible in our trEPR experiment. We measure a frozen solution sample where we do not expect large-scale conformational dynamics but only vibrations and small-scale torsional motion. We link the dependence on excitation energy instead to the static distribution of conformers (Figure 1D) and changing character of the initial state. Long-time spin evolution is favored in less coupled conformations and when the excitation targets the higher excited states. The differences in frozen state trEPR with excitation energy suggests that further mechanistic detail could be obtained by routinely investigating conformational heterogeneity in dimer samples.

## Discussion

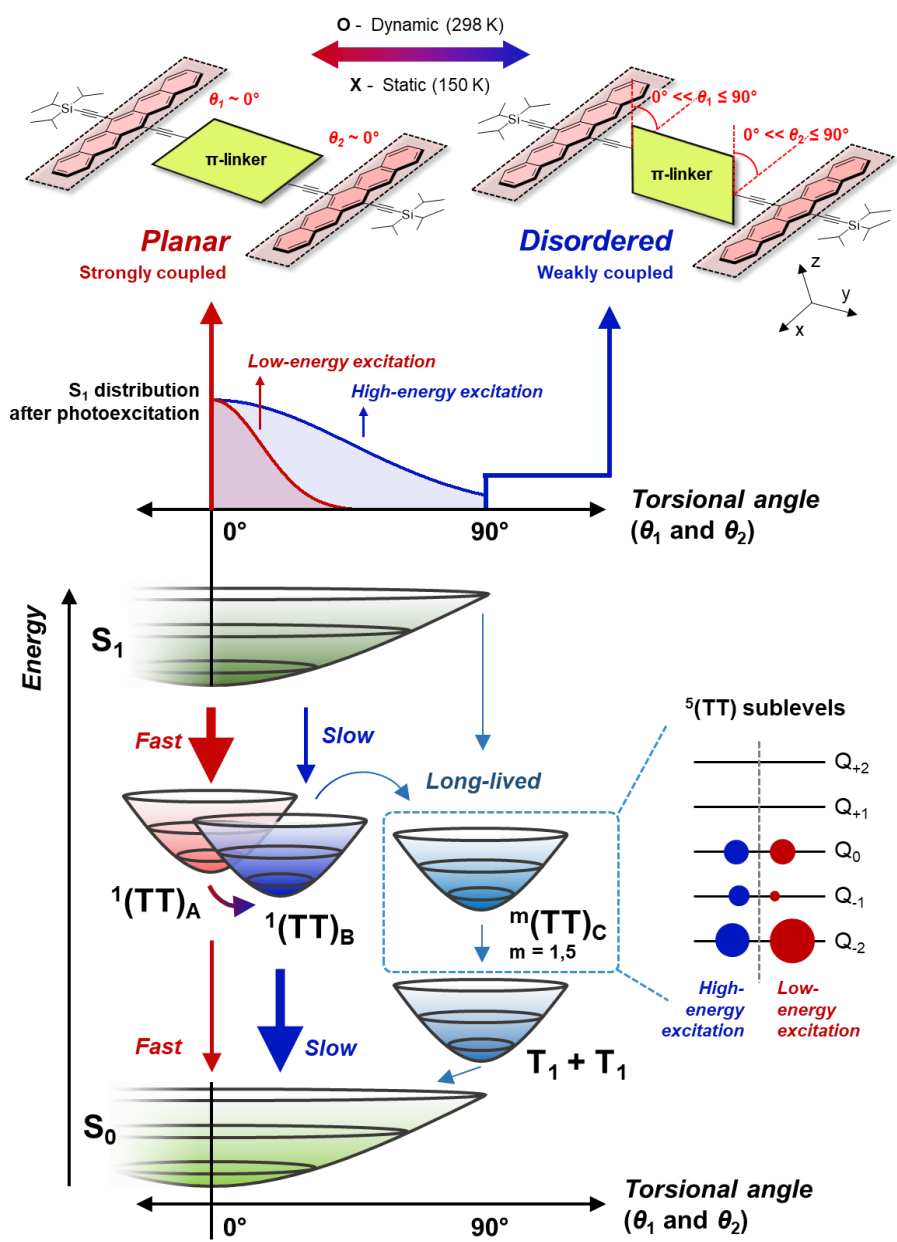

**Figure 7.** Schematic illustration of the heterogeneous singlet fission pathways in 2P-PDPP.

**Generality of Heterogeneous Singlet Fission.** Our systematic excitation-energy-dependent TA and trEPR results for 2P-PDPP highlight that measurement with a single excitation wavelength is inadequate to fully characterize the SF behavior, nor can a single ensemble with well-defined rate constants describe our observations. The TT formation and recombination dynamics, involvement of multiple intermediates, and the nature and balance of long-lived states all hinge on an interplay between static and dynamic heterogeneity. Only by evaluating the response of distinct, selectively excited sub-ensembles can we capture a unified picture of the complex potential energy landscapes that govern SF in 2P-PDPP (Figure 7). While we describe this behavior specifically in the context of 2P-PDPP, the underlying principles behind conformational heterogeneity and its impact on interchromophore coupling are very general. Thus, this behavior should apply widely to intramolecular SF systems, and we consider that this dynamic picture of heterogeneous SF should supplant the more generally applied static model.

A key criterion for this structural heterogeneity to have an impact is a sufficiently fast rate of SF. In typical systems, vibrational relaxation/cooling or solvation within $S_1$ will drive most conformers to the same equilibrium geometry. These processes typically occur on ~ ps timescales, in exceptional cases into the 10's of ps regime,[22] setting a critical threshold for the onset of conformational effects. In tetracene dimers, TT formation is frequently endothermic or isoergic, and it accordingly proceeds over ~ ns timescales where heterogeneity would be unanticipated. On the other hand, a small set of tetracene dimers with strong through-bond coupling reveal much faster TT formation,[37,51] in direct competition with typical vibrational dynamics. In these, we would expect heterogeneous SF to play out through a complex interplay between energetics and $S_1$-TT couplings, though this was not addressed in prior reports. At another extreme, the excess exothermicity of TT formation in hexacene dimers generally results in slow (few-ps timescale) SF,[51] like electron transfer reactions in the Marcus-inverted region. This timescale is moderately competitive with thermalization, and these materials could likewise reveal heterogeneous SF. The most widely studied SF dimers, though, are composed of pentacenes. With SF energetics ranging from isoergic to moderately exothermic, and widespread reports of sub-ps SF, these model systems are easily the most susceptible to the effects of heterogeneity.

A review of the rich library of reported pentacene dimers suggests that the type of linker and bridging geometry, which controls the interchromophore coupling strength,[17] are likewise determining factors for heterogeneous SF. In the widely studied 2,2'-linked P-TIPS dimer family, phenyl and related bridges can provide some degree of conjugation between pentacenes,[15] but the combination of steric hindrance and the strong local aromaticity of benzene makes this a relatively minor effect. Hence, the steady-state absorption spectra of most 2,2'-dimers are nearly identical to monomeric pentacenes. They exhibit relatively slow SF, where excess energy in $S_1$ is dissipated to the surroundings well in advance of TT formation, and we anticipate the excitation energy would have little impact. A key exception is the directly linked (i.e. bridge-free) 2,2'-dimer, where the dihedral angle has been shown to dramatically affect the coupling strength between $S_1$ and $^1$(TT).[52] In this system, slight heterogeneity in the inter-pentacene angle should yield significant variation in the initial SF rate. A similar effect is observed in the equivalent directly linked pentacene dimers with 6,6' connectivity.[13,41] In polar media, red-edge excitation of these molecules selectively populates a relatively planarized (i.e. more strongly coupled) $S_1$ state, with the subsequent triplet yield strongly dependent on the relative energy of nearby CT states. Still, in such directly linked dimers, steric hindrance results in steep dihedral angle potentials, meaning that these strongly coupled sub-ensembles are very much a minority population. This limitation is lifted in 6,6'-dimers where the bridge incorporates acetylene linkers,[37,38] a common motif that includes 2P-PDPP. These moieties simultaneously permit unobstructed dihedral rotation and provide effective conjugation even in torsionally distorted conformations. A hallmark of the enhanced electronic (or vibronic) coupling within the dimers is considerable deviation of their steady-state absorption spectra compared to their monomeric counterparts, as shown in Fig. 1. This combination of strong coupling (i.e. rapid SF) and minimal barriers to distortion makes such structures particularly prone to the effects reported here. To meaningfully connect the molecular structure to photophysical function, it is essential to evaluate the behavior of selectively excited sub-ensembles, with especial emphasis on the red edge excitation to probe the most stable ground-state conformations.

As shown in the foregoing, our combined time-resolved electronic and spin resonance data for 2P-PDPP over a range of excitation energies reveals that the choice of conformer impacts the timescales of TT formation and recombination, the magnitude of long-lived population, and even the balance of spin sublevels, that is, the full set of important SF characteristics. Interestingly, these ensembles are not static, and the SF kinetics are strongly affected by conformational dynamics (Figures 3D-F). As noted above, we observe that the relative balance of 'fast' channels (characteristic of planar, type-A geometries) and 'slow' channels (representative of more twisted B-type structure) reverses between the TT formation and recombination timescales. This behavior points to structural relaxation within TT that reduces the conjugation and thus spin-spin exchange coupling that governs the long-time dynamics of TT. Such structural changes are contrary to what is typically observed within $S_1$ states, where relaxation tends to extend exciton delocalization and increase the interactions between chromophores, but they are consistent with the favorable entropic effects of a distorted TT state.[53,54] Entropy alone provides insufficient driving force, though, since the $^1(TT)_A \rightarrow {}^1(TT)_B$ conversion is suppressed near band-edge excitation. This excitation energy dependence points to the presence of an activation barrier. We recall that the initial SF dynamics outcompete thermalization within $S_1$, meaning that the $^1(TT)$ state formed is vibrationally hot. There is thus ample excess energy to drive $^1(TT)_A$ to $^1(TT)_B$ conversion in competition with energy dissipation to the environment, especially following excess-energy excitation. The persistence of this process at lower efficiency even following band-edge excitation indicates that the relevant torsional barriers are relatively shallow. These barriers reveal an unexpected way that excitation energy can steer relaxation processes within TT and help determine overall triplet yields.

**Origin of Quintet Sublevel Population Variation.** However, it remains challenging within this framework to fully explain the excitation behavior of long-lived $^1(TT)_C$ states. Whereas the TT formation/recombination effects largely saturate for increasing excitation energies beyond 1.9 eV, the yield of $^1(TT)_C$ monotonically increases through the entire measured range (Figure 3F). The 1 eV energy range spanned dwarfs the torsional barrier (~45 meV) and is well beyond the range where we would anticipate marked differences in absorption spectra between 2P-PDPP conformers. Instead, we propose a complementary effect due to the changing character of the initial excitation. In the relevant spectral range, the excitation is increasingly resonant with the PDPP bridge and higher-lying pentacene excited states, both of which are associated with enhanced CT character. We infer that the presence of such CT character in the initial excitation increases the SF yield in the distorted geometry. While the detailed mechanism of this effect is not clear, it is consistent with the similarly excitation-dependent CT signature observed in trEPR (Figure 6A).

Indeed, the quintet and triplet signals in trEPR, as well as the trends in sublevel populations, follow a strikingly similar energy dependence, indicating that they are governed by a comparable interplay between static and dynamic heterogeneity, including a prominent role for the changing character of higher-energy excited states. We consider it remarkable, though, to observe such similar behavior in such fundamentally different samples. The TA measurements are performed in room-temperature solution, where large-scale conformational dynamics face little barrier. Not only do the frozen solutions used for trEPR drastically limit such conformational change, but we obtain quintet signal only for dimers with (near) coplanar pentacenes. In this limited sub-ensemble, through-space pentacene-pentacene coupling is maximized, while the (potentially stronger) through-bond interactions are substantially modulated by the 'twist' of the PDPP-Hex bridge. We expect these conditions are particularly favorable for long-time spin evolution. As we discuss below, the same interplay between static heterogeneity and structural dynamics in this restricted conformational space gives rise to our novel observation of ESP tuning through excitation energy.

Our trEPR study has shown that quintet state sublevel populations are strongly dependent on the excitation energy while the free triplet sublevel populations are rather insensitive. The quintet state spectra of 2P-PDPP show a decrease in $Q_{-2}$ and increase in $Q_{-1}$ sublevel populations together with an increasingly negative $S$ parameter when utilizing higher excitation energies. To shed some light on the mechanistic differences that give rise to our complex observations, we look to the study by Collins *et al.*, which showed that population of the quintet sublevels is strongly dependent on the time dependent exchange coupling $J$.[55] This parameter is considered to change from $|J| \leq |D|$ to $|J| \gg |D|$ during SF and subsequent spin evolution. The change in exchange coupling strength is associated with nuclear reorganization within the molecule(s), and in the case of dimers usually torsion of the bridge moiety plays a crucial role. Even at the low temperatures conventionally used for trEPR measurements this torsional motion is sufficiently active. Collins *et al.* suggested three parameters that determine the quintet sublevel populations within the time dependent exchange coupling model.[55] The first parameter, denoted $t_0$, is the delay or activation time before the change in exchange coupling and nuclear reorganization occurs. $t_0$ is non-zero when the excited state potential surface exhibits an energy barrier to nuclear reorganization. The second parameter is $\tau_s$, which is the statistical lifetime. $\tau_s$ governs the distribution in activation times $t_0$ and gives the probability distribution $P = \frac{1}{\tau_s} e^{-\frac{t_0}{\tau_s}}$. The last parameter is $\tau_r$, the time taken for the exchange coupling to rise from $J_{min}$ to $J_{max}$ via the relationship $J = J_{max}(1 - e^{-(t-t_0)/\tau_r})$ for times $t > t_0$. According to this framework, following fast $\tau_r$ (e.g. 100 fs) the quintet population will be predominantly in the $Q_0$ sublevel, while slow $\tau_r$ (e.g. 100 ps—10 ns) yields quintet population chiefly in the $Q_{-2}$ and $Q_{-1}$ sublevels.[55] In our case, simulations that do not account for the time-dependent exchange coupling suggest that quintet population is predominantly in the $Q_{-2}$ and $Q_0$ sublevels.

Quintet sublevel population residing predominately in $Q_{-2}$ and $Q_0$ is not unusual and was similarly observed for pentacene aggregates in non-glass forming frozen solution.[56] In this case too, the quintet sublevel population was attributed to a change in the exchange coupling during the SF process, this time from a strongly coupled pair with $J_{max}$ to a weakly coupled pair with $J_{min}$. The time-dependent exchange coupling was attributed to a change in the TT pair distance occurring due to exciton hopping in a disordered frozen matrix with different aggregate environments. While our trEPR measurements are also in a non-glass forming frozen solution, we rule out aggregation leading to the observed excitation dependent quintet spectra based on the excitation-energy-independent quintet and triplet kinetics (Figure S23) and relatively unchanging triplet/quintet ratio (Figure S24). If excitation energy was selecting different aggregate sizes, we would expect differences in the kinetics and the triplet/quintet ratio.[57] Furthermore, a similar quintet spectrum to ours has been observed in a pentacene dimer system in a glass-forming frozen solvent.[40] Hence, we attribute the quintet excitation-energy-dependent spectra to arise from the mechanism described by Collins *et al.* whereby torsional motion results in a time-dependent exchange coupling which, depending on $t_0$, $\tau_s$, and $\tau_r$, can lead to the observed quintet sublevel populations.[55] It is important to note that for dimer systems the change in exchange coupling during SF can be from $J_{min}$ to $J_{max}$ or vice versa, depending on the ground- and excited-state geometries. Here, we identify varying through-bond coupling due to fluctuations in the bridge dihedral angle as the most likely coordinate.

**Tuning of Ordering Parameters and Relative Yields of Quintet and Free Triplets.** Next, we look to qualitatively explain the increasingly negative $S$ (ordering parameters) needed to model quintet spectra with increasing excitation energies. In the limit of $|J| \leq |D|$, dipolar coupling leads to spin mixing of the $^1(TT)_0$ and $^5(TT)_0$ states to form mixed states, denoted SQ and QS. During the period until $t_0$ the singlet projection onto the SQ and QS state is time-dependent with oscillation periods determined by the ZFS parameters. When the external magnetic field is parallel to the $X,Y$ orientation, the period of oscillation is approximately given by $2\pi/(D_T \pm 3E_T) \sim 5.5$ ns and in the $Z$ orientation by $\pi/D_T \sim 2.78$ ns.[55] At times after $t_0$ the singlet projection no longer oscillates and due to the different oscillation periods at $X,Y$ orientations compared to $Z$ orientations this leads to different probabilities of $^5(TT)_0$ populations at the different molecular orientations. Therefore, the reason for the observed increasingly negative $S$ with higher excitation energies must be due to $t_0$ changing such that we move closer to $t_0 = n \times \frac{5.5\text{ns}}{2}$ (where $n = 1,3,5 \ldots$) and further from $t_0 = m \times 5.5$ ns ($m = 1,2,3 \ldots$). Since the oscillation period at the $X,Y$ orientation is roughly half that of the $Z$ orientation, at $n \times \frac{5.5\text{ns}}{2}$ the maximum intensity of the $X,Y$ orientation would be observed relative to the $Z$ orientation. As $t_0$ (activation time for nuclear reorganization) is related to the excited state potential surface, we propose that higher energy excitations lead to higher vibrationally excited manifolds resulting in different $t_0$ values.

Lastly, we come to the increased relative yields of quintet and free triplets with higher excitation energies. While we observe quintet states only from dimers with co-planar pentacenes, as shown by the DFT results even at 150 K a large range of dihedral angles between aP-TIPS and PDPP-Hex can still be occupied. Hence, we expect that bridge heterogeneity (starting geometry) along with its torsional motion (degree of motion accessible) would lead to a wide variety of excited state potential surfaces which would have differing degrees of SF quintet and free triplet yields, based on the time-dependent exchange coupling.

While the above discussion only gives qualitative reasoning for the parameters that would govern the quintet state spectra for 2P-PDPP as a function of different excitation energies, routinely applying such trEPR studies on dimer systems would allow for better understanding of how these parameters differ based on molecular structure, linkers/linker positions and solvent dielectrics. One key question is whether static heterogeneity in frozen solutions or the higher excited states surfaces accessed through higher excitation energies play the decisive role. In our case, the linker is itself a chromophore, and hence the behavior of this dimer could indeed be different from other dimer systems, highlighting the need for more excitation energy dependent studies on SF systems. Further work to quantitatively characterize the time dependent exchange couplings will form the basis of future work.

## Conclusions

In conclusion, we turn to the broader question of how such dimers function as model systems for the complex photophysics of thin films. In general, thin films are structurally 'static', but their excited-state processes are decidedly richer than the fixed structure suggests. They are typically 'dynamic' systems, characterized by exciton migration or relaxation through a broad density of states that brings with it changes in state energies and couplings. This density of states can be modulated—tuning from amorphous films to polycrystalline films to single crystals—but even in crystals, the presence of defects and surface states yields subtle differences in molecular environment. These effects collectively contribute to the great challenge of establishing robust structure-property relationships in the solid. This has provided impetus for the study of synthetically tailored molecular dimers, which can be viewed as the smallest repeating unit in thin films. They are typically treated as representative of one of many possible 'static' conformations to understand a cross-section of the complex 'dynamic' relaxation pathway. Our excitation-dependent TA and trEPR results, however, indicate that this approach is oversimplified, since dimers can also be governed by heterogeneous relaxation arising from dynamically varying and intrinsically varied interchromophore coupling strength. That is, they can be just as 'dynamic', sampling a wide distribution of conformations and couplings through their relaxation. They may thus prove to be better than expected models for the solid state, able to capture processes that depend on the broad density states like entropic effects. However, this interpretation hinges on the analogy between a single, highly dynamic molecule, and a disordered array of static molecules, which could behave fundamentally differently. Consequentially, we suggest that some caution is in order for treating dimer complexes as well-defined model systems for thin films.

## Acknowledgements

The DFT calculations were supported by the National Institute of Supercomputing and Network (NISN)/Korea Institute of Science and Technology Information (KISTI) with needed supercomputing resources, including technical support (KSC-2022-CRE-0428). W.K. was supported by the National Research Foundation of Korea (NRF) grant funded by the Korean government (MSIT) (RS-2023-00210400). N.A.P and R.B. acknowledge funding by the Deutsche Forschungsgemeinschaft (DFG, German Research Foundation) under Germany's Excellence Strategy—EXC 2008—390540038—UniSysCat. S.P. thanks Science and Engineering Research Board for funding the project through the IRHPA grant IPA/2020/000033 and core research grant CRG/2022/004523. A.J.M. acknowledges partial support of this work through a Cornell University College of Arts and Sciences New Frontiers Grant.

## Author Contributions

S.P. and A.J.M. conceived the project. K.M. synthesized the sample under the supervision of S.P.. W.K. performed DFT calculations and TA experiments, and analyzed the results with A.J.M.. N.A.P. performed trEPR experiments and analyzed the results with R.B.. All authors discussed and commented the experiments and results. W.K., N.A.P., and A.J.M. wrote the paper with input from all authors.

## Competing Interests

The authors declare no competing interests.